\documentclass
[aps,prb,twocolumn,groupedaddress,showpacs]{revtex4}%
\usepackage{epsfig}
\usepackage{amsmath}
\usepackage{amsfonts}
\usepackage{amssymb}%
\providecommand{\U}[1]{\protect\rule{.1in}{.1in}}
\begin{document}
\title{Quasiparticle interference in antiferromagnetic parent compounds of Fe-based superconductors.}
\author{I.I. Mazin$^{1}$, Simon A.J. Kimber$^{2}$, and Dimitri N. Argyriou$^{2} $}
\affiliation{$^{1}$Code 6393, Naval Research Laboratory, Washington, DC 20375, USA}
\affiliation{$^{2}$ Helmholtz-Zentrum Berlin f\"{u}r Materialien und Energie, Hahn-Meitner
Platz 1, Berlin 14109, Germany}

\begin{abstract}
Recently reported quasiparticle interference imaging in underdoped
Ca(Fe$_{1-x}$Co$_{x}$)$_{2}$As$_{2}$ shows pronounced C$_{2}$ asymmetry that
is interpreted as an indication of an electronic nematic phase with a
unidirectional electron band, dispersive predominantly along the $b$-axis of
this orthorhombic material. On the other hand, even more recent transport
measurements on untwinned samples show near isotropy of the resistivity in the
$ab$ plane, with slightly larger conductivity along $a$ (and not $b$). We show
that in fact both sets of data are consistent with the calculated ab initio
Fermi surfaces, which has a decisevly broken C$_{4}$, and yet similar Fermi
velocity in both directions. This reconciles completely the apparent
contradiction between the conclusions of the STM and the transport experiments.

\end{abstract}

\maketitle

The Fe-based superconductors present a new paradigm for high-$T_{C}$
superconductivity as here Cooper-pairs appear to emerge upon chemical doping
from a metallic ground state as opposed from a Mott insulator as found in the
celebrated High-$T_{C}$ cuprates\cite{Lee}. Despite this difference of parent
ground state of the Fe- and Cu-based superconductors, similarities lie in that
in both cases superconductivity emerges after the suppression of static
ordered magnetism\cite{I}. Although band theory has correctly predicted the
unusual antiferromagnetic (AFM) order in the parent compounds of the Fe-based
superconductors, it consistently overestimates the tendency to magnetism and
underestimates the electronic mass, so there is no doubt that electronic
interactions can not be ignored in quantitative descriptions, and that they
play a different role compared to cuprates. The exact role of correlations,
especially once the parent phase of the Fe-superconductors is doped, has been
the focus of much debate and controversy.

An almost universal feature of the Fe-superconductors is that in the parent
phases, there is a tetragonal to orthorhombic structural phase transition that
is closely associated with the onset of antiferromagnetic order\cite{Review}.
Upon chemical doping $x$, the onset of the structural and magnetic transitions
($T_{S}$ and $T_{N}$ respectively) decrease with $x$ and superconductivity
emerges. The physical nature of the cross over from antiferromagnetic order to
superconductivity varies between specific materials. In some cases both
$T_{S}$ and $T_{N}$ coincide while in others $T_{S}$ is a few degrees higher
than $T_{N}$\cite{Review}.

Band structure calculations have suggested that the AFM ordering is
accompanied by a strong restructuring of the Fermi surface, with the Fermi
surface area being reduced by roughly an order of magnitude. This has been
confirmed by optical and Hall measurements that register a drastic reduction
of the carrier concentration in the AFM state\cite{Opt_Hall}. The calculated
AFM Fermi surface consists of several small pockets, which are arranged in the
Brillouin zone in a way that strongly breaks the tetragonal symmetry, but each
of them is rather isotropic\cite{8}. This led to a prediction of small
transport anisotropy. An alternative point of view, that associates the
orthorhombic transition with orbital (charge) degrees of freedom, suggests a
double exchange (metallic) ferromagnetic interaction along one
crystallographic direction and a superexchange along the other direction. This
picture is also consistent with the observed AFM order and naturally suggests
a metallic conductivity along the ferromagnetic chains and a substantially
reduced conductivity in the other direction.

Recent experiments on detwinned single crystals support the former point of
view: they demonstrate a small anisotropy with the AFM direction being $more,$
not $less$ metallic. However, transport measurments are integrated probes, and
also involve possibly anisotropic scattering rate, therefore experiments
directly probing the topology of the Fermi surface in the AFM state are highly desirable.

One such experiment has been recently performed by Chuang \textit{et
al.}\cite{1}. They have reported quasiparticle interference (QPI) imaging of a
lightly cobalt doped sample of CaFe$_{2}$As$_{2}$ compound. They interpreted
their result in terms of a quasi-1D (\textquotedblleft
unidirectional\textquotedblright) electronic structure, metallic only along
the FM, consistent with above-mentioned orbital picture. On the the other
hand, their argumentation was rather indirect, based largly on the fact that
directly measured dispersion of the QPI maxima (which was indeed 1D) coincded
with the ARPES-measured band dispersion along the the same direction.

In this paper we show that in reality the data of Ref. \onlinecite{1} are
consistent with the calculated ab initio Fermi surfaces, and not with the
implied in that work 1D bands. This reconciles completely the apparent
contradiction between the conclusions of Ref. \onlinecite{1} and the transport
measurements on untwinned samples.

The reported STM examination shows a QPI pattern in the momentum space that
breaks completely the $C_{4}$ symmetry, the main features being two bright
spots along the $y$ (crystallographic $b)$ direction, with no counterparts
along $x$ (note that $y$ is the $ferromagnetic$ direction, and $x$ in the
\textit{antiferromagnetic} one). Ref. \onlinecite{1} insists \textquotedblleft
that the scattering interference modulations are strongly unidirectional,
which should occur if the k-space band supporting them is
nematic\textquotedblright. However it should be kept in mind that this occurs
in that part of the phase diagram where the long-range antiferromagnetic order
is fully established, as reflected by the fact that the lattice symmetry is
orthorhombic, and the $C_{2}$ symmetry is already completely broken. Indeed
the size of the orthorhombic distortion is not \textquotedblleft
minute\textquotedblright, as Ref. \onlinecite{1} posits, with $b/a$ $\sim1$\%,
and is instead comparable with distortions seen in various iron oxides
systems. For instance, in the Verwey transition the Fe-O bond dilation is
$\sim$0.6\% with Fe atoms in the same tetrahedral symmetry as in the
ferropnictide superconductors{\cite{6}}, and this is usually considered to be
a strong distortion. Similarly, in the antiferromagnetic phase of FeO, where
the cubic symmetry is completely broken, the structural effect is also on the
same order\cite{7}.

Since the sample under study is orthorhombic it is misleading to call its
electronic structure nematic, as the lattice orthorhombic distortion here is
substantial. Nematic phases are frequently found in organic matter. The
defining characteristic of these phases is orientational order in the absence
of long range positional order, resulting in distinctive uniaxial physical
properties. It has also been proposed that nematic order exists in some
electronic systems, and may even play a role in mediating high temperature
superconductivity\cite{4}. Borzi \textit{et al}\cite{5} demonstrated the
presence of another interesting phase in Sr$_{3}$Ru$_{2}$O$_{7}$ at
millikelvin temperatures and high magnetic fields, which has also been called
nematic. In this case, the crystallographic planes were shown to remain
strictly tetragonal (withing 0.01\%) with $C_{4}$ structural symmetry, while a
pronounced $C_{2}$ asymmetry in electronic properties was measured. This
breaking of the electronic symmetry compared to that of the underlying lattice
is now conventionally referred to as electronic nematicity (in fact, even in
those cases one has to be careful to distinguish between nematic physics and
simply an unusually weak electron-lattice coupling, but this goes beyond the
scope of this paper, and in any event is not a concern for Fe pnictides where
this coupling is strong).

Since the tetragonal symmetry is decisively broken at the onset of the
magnetic order in this ferropnictide, it is clear that the symmetry of the
electronic structure defining the structural distortion is also completely
broken. What is more important is that while the observed QPI pattern does
violate the $C_{4}$ symmetry, it is clearly not one-dimensional, in the sense
that it varies equally strongly along $k_{x}$ and $k_{y}$ directions. Thus,
interpretation of the data in terms of a 1D electron band does not appear to
be possible. To understand this experiment one needs to start with a realistic
model for the electronic structure and actually calculate the QPI pattern.

Such calculation has recently been presented by Knolle $et$ $al$\cite{11}.
They used a weak-coupling theory that interprets tha antiferromagnetic state
as resulting from a spin-Peierls transition, with a correspondingly small
magnetic moment. Knolle $et$ $al$ have been able to describe qualitatively the
experimental data obtained by Chuang $et$ $al$ in the sense that their
calculated QPI pattern strongly breaks the $C_{2}$ symmetry, while the band
dispersion, on average, remains fairly isotropic in plane. Note that one
should not be looking for a \textit{quantitative} interpretation, since the
STM experiment in question did not detect any Ca atoms on the surface, so the
sample surface is likely charged with up to 0.5 hole per Fe, and thus any bulk
calculation can only be applied to this experiment in a qualitative way.
Besides, it was recently shown\cite{SS} that Fe pnictide systems feature
surface states quite different from the bulk that should undoubtedly affect
the STM spectra.

However, this result, as mentioned, has been obtained in a weak coupling
limit, corresponding to small magnetization, while in this system the ordered
magnetic moments are on the order of 1 $\mu_{B},$ and local moments even
larger\cite{2,local,12}. Not surprisingly, their Fermi surface is rather far from
that measure recently on untwinned samples by Wang et al\cite{Dessau}, while
the LDA Fermi surface reproduces it quite well\cite{note}. Indeed, this is a
known problem in the weak coupling approach: while being physically justified
for the paramagnetic parts of the phase diagram, the Fe magnetism in the
ordered phases is driven by the strong local Hund rule coupling, and not by
the Fermi surface nesting, as assumed in the weak copling models.

Therefore we have calculated the QPI images for antiferromagnetic CaFe$_{2}%
$As$_{2}$ entirely from first principles\cite{note2}, using the Local Density
Approximation (LDA) magnetic moment (somewhat larger that the experimental
moment at zero doping). We used the standard linear augmmented plane wave
method as implemented in the WIEN2k code\cite{W2k}. The corresponding Fermi
surface is shown in Fig. 1. We see that the magnetism has a drastic effect on
the Fermiology, and the resulting Fermi sirfaces are completely breaking the
$C_{4}$ symmetry. Apart from small quasi-2D tubular pockets, originating from
Dirac cones, there is one hole pocket around Z (0,0,$\pi/c$ or 2$\pi/a$,0,0)
and two electron pockets between Z and 0,$\pi/b$,$\pi/c$. It is immediately
obvious that the QP scattering between these pockets must exhibit strong
interference for scattering along $b,$ but not $a.$

\begin{figure}[t]
\includegraphics[width=0.8\columnwidth]{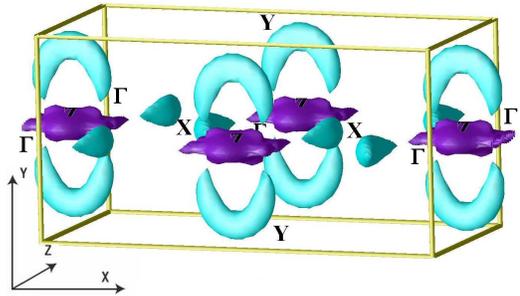}\caption{(Color
online) Calculated LDA Fermi surface for CaFe$_{2}$As$_{2}$ in the
antiferromagnetic state.}%
\label{Fig:1}%
\end{figure}\begin{figure}[t]
\includegraphics[width=0.8\columnwidth]{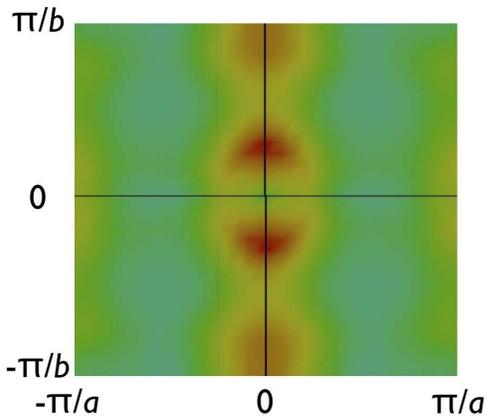}\caption{(Color
online) Quasiparticle interference pattern (in arbitrary units) for zero bias and q$_{z}$$\sim$0,
calculated using the same electronic structure as in Fig. \ref{Fig:1} and Eq.
1.}%
\label{Fig:2}%
\end{figure}

Indeed, we have calculated the QPI function $Z,$ using the known expression
(Ref. \onlinecite{9}, Eq. S9)%
\begin{equation}
|Z(\mathbf{q},E^{\prime}\mathbf{)|}^{2}\mathbf{\propto}\int\frac{dE^{\prime}%
}{E-E^{\prime}}\sum_{\mathbf{k}}\delta(E-E_{\mathbf{k}})\delta(E^{\prime
}-E_{\mathbf{k+q}}),
\end{equation}
where we assumed a constant inpurity scattering rate and a constant tunneling
matrix elements. This approximation is sufficient for a qualitative or
semiquantitative comparison. As explained above, given that the surface in the
experiment in question was charged compared to the bulk, a quantitative
comparison is meaningless.

A calculated pattern (there is some dependence on $q_{z}$ and on $E,$ but we
are interested in the qualitative features only) are shown in Fig.2. One can
see iimediately that, very similarly to the patterns obtained in Ref.
\onlinecite{1}, two sharp maxima appear at $\mathbf{q}=0,\pm\xi,0,$ where
$\xi\sim\pi/4b$. The origin of these QPI features is obvious from the Fermi
surface (Fig. 1). Note that these LDA calculations have no adjustable
parameters, and yet are in excellent qualitative agreement with the QPI images.

It is also worth noting that while the calculated Fermi surfaces completely
break the tetragonal symmetry, which is fully reflected in the QPI images, the
individual pockets are very three-dimensional, so that the calculated
conductivity is comparable for all three directions\cite{8}. While
experimentally there is up to a 20\% $a/b$ charge transport anisotropy\cite{8}
close to tetragonal to orthorhombic phase boundary in CaFe$_{2}$As$_{2}$, it
is much less than what would be predicted for a quasi 1D electronic band, and
of the opposite sign\cite{10}.

It may be worth at this point to explain at some length while a quantitative
comparison between a Fourier transform of a tunneling current map, and
theoretical calculations, whether ours or any other, is impossible at this
stage. Quasiparticle interference, as discussed in many papers, manifests
itself in tunneling in a very indirect way. In a sense, it is a multistage
process. First, a defect existing near the metal surface, is sdreened by the
conducting electrons. This creates Friedel oscillations in the real space.
This oscillations are formed by all electrons (mostly those near the Fermi
surface, but not only). In a multiband system, it includes electrons
originated from different atomic orbitals, such as $xy,$ $xz,$ $yz,$ $z^{2}$
and $x^{2}-y^{2}.$ As is well known in the theory of tunelling, the rate at
which electrons tunnel through vacuum depends drastically on their orbital
symmetry, especially on their parity (see, e.g., Ref. \cite{EPL}). Indeed
tunelling through a wide barrier mainly proceeds through electrons with zero
momentum projection onto the interface plane (such electrons have to travel
the shortest lengths in the subbarrier regime). If such electrons belong to an
odd 2D representation (for d-electrons, all but $z^{2},$ if $z$ is the normal
direction), the tunneling rate is suppressed. This effect is well known in
spintronics, where it can drastically change the current spin polarization. On
the other hand, for a thin barrier the tunneling conductance depends on the
number of the conductivity channels, which is given by the density of states
(DOS) times normal velocity. In both cases, it is not just the density of
quasiparticles, as assumed in Eq. 1 (and in Ref. \cite{11}), but the DOS
weighed by a strongly \textbf{k}-dependent, unknown function. 

Nothing is known about the nature of the scattering centers, producing the
above mentioned Friedel oscillations. In this particular experiment they may
be magnetic or nonmagnetic defects, twin domain boundaries, antiphase domain
boundaries, remaining surface Ca ions, and more. Some of these scatterers are
strongly anisotropic by nature, others are strongly dependent on the orbital
character. We have dropped the scattering matrix elements completely form our
consideration. Knolle $et$ $al$\cite{11} instead have chosen a specific model
for the scattering centers. We believe that without any knowledge about the
actual scattering centers in the system any QPI using a particular model is
more obscuring the actual physics, compared to the simplest constant matrix
elements approximation, rather than clarifying it.

Finally, there are several issues specific for this particular experiment: (1)
unknown, but strongly different from the bulk, charge state. As opposed to
Ba122, and Sr122, where 1/2 of the alkaline earth atoms stay on the surface,
providing charge neutrality, in Ca122 STM does not detect any Ca on the
surface, suggesting a strongly charged surface. A corollary of that is
appearence of a surface reconstruction (as indeed observed), of a surface
relaxation, and, importantly (since tunneling  proceeds largely through the
surface states), of surface bands (as demonstrated, for instance,
in Ref. \cite{sb}.

While the above considerations preclude a quantitative comparison and extracting
quantitative analysis of the experiment in question, we see, particularly
when comparing our calculations with those of Knolle $et$ $al$\cite{11}, 
that the $C_{2}$ QPI structure observed in Ref. \onlinecite{1} is a very
universal consequence of the long-range stripe-type antiferromagnetic
ordering. Indeed, Knolle $et$ $al$ calculations were built upon a besically incorrect
band structure and fermi surfaces, an used a weak coupling nesting scenario for the
antiferromagnetism, while in reality the magnetism in pnictides is a strong coupling phenomenos;
yet, their calculations produced a ``unidirectional'' QPI pattern just as well.
Together with the strong-coupling LDA
calculations, this span a large range of possible models, indicating that 
the $C_{4}$ symmetry is strongly broken in QPI images with  simply by virtue
of the long range AFM order, whatever the the origin of this order. 

Last but not least, we can also predict, from our calculations, that
this symmetry will be also broken, although the peaks are likely to be
substantially broaden, in the truly $nematic$ phase (see review \cite{2} for a
discussion), that is to say, the phase between the long-range magnetic
transition and the structural orthorhombic transition.


\begin{thebibliography}{99}
\bibitem {note}Compared to Ref. \cite{Dessau}, both $\beta$1 and $\beta$2
bands are present, their nontrivial crescent shape is reproduced, their size
and location (around the Z point) are consistent with the calculation. Not
that these $\beta$ pockets are mainly responsible for the QPI peak in our Fig.
2. The flattish $\gamma$ pocket is also in excellent agreement with the
calculation, although in the experiment it is split into $\gamma$3 and
$\gamma$4 (probably an effect of the surface reconstruction). The claimed
experimental bands ($\alpha$1, $\alpha$2, $\gamma$1 and $\gamma$2) along Z-X
are quite messy. The calculations predict small pockets there located roughly
where ARPES sees some bands. These are formed by the famous \textquotedblleft
Dirac cones\textquotedblright. The only feature that does not find any
correspondence in the calculation is the long segment \textquotedblleft%
$\gamma$2\textquotedblright\ stretched along $k_{y}$. This may be a surface
state similar to those discovered in Ref. \onlinecite{SS}(note that this band
is drawn rather speculatively, the corresponding signal is really weak). This
agreement is even more impressive given that the Fermi surface reproduced here
was published 7 months ago (Ref. \onlinecite{8}), well before any untwinned
ARPES data became known.



from 1D bands.
\end{thebibliography}
\end{document}